\documentstyle[multicol,prb,aps,epsf]{revtex}

\begin{document}
\title{Generalized boundary conditions for the circuit theory of spin transport }
\author{Daniel Huertas-Hernando$^{1}$, Yu. V. Nazarov$^{1}$ and W. Belzig$^{2}$}
\address{$^{1}$Department of Applied Physics and Delft Institute of Microelectronics\\
and Submicrontechnology, \\
Delft University of Technology, Lorentzweg 1, 2628 CJ Delft, The Netherlands.%
\\
$^{2}$ Department of Physics and Astronomy, University of Basel,\\
Klingelbergstr. 82, CH-4056 Basel, Zwitzerland. }
\date{\today}
\maketitle

\begin{abstract}
The circuit theory of mesoscopic transport provides a unified framework of
spin-dependent {\em or} superconductivity-related phenomena. We extend this
theory to hybrid systems of normal metals, ferromagnets {\em and}
superconductors. Our main results is an expression of the current through an
arbitrary contact between two general isotropic nodes. In certain cases
(weak ferromagnet and magnetic tunnel junction) we derive transparent and
simple results for transport properties. 
\end{abstract}

\begin{multicols}{2}%
%

\section{INTRODUCTION}

Spin-transport in hybrid ferromagnet (F)-normal metal (N) systems has been
object of extensive investigation since the discovery of the giant
magnetoresistance (GMR) effect.\cite{RefLevy} Magneto-electronic
multiterminal systems lead to novel applications, {\em e.~g.} magneto
recording heads or magnetic-field based sensor devices. Future prospectives
of non-volatile electronics motivate also many fundamental and applied
research. Transistor-like effects were found in a ferromagnet-normal metal
three-terminal device, that depend on the relative orientation of the
magnetization of the ferromagnets.\cite{RefMJohnson,RefJohnsontrans} The
dependence of transport properties with the relative angle between the
magnetization directions of the ferromagnets, was addressed experimentally 
\cite{RefMoodera1} as well as theoretically\cite{RefUstinov} in the past.
These days non-collinear spin transport attracts an increased attention due
to the recent interest on the spin-current induced magnetization torques. 
\cite{RefSlonczewski,RefKatine,RefWaintal} A ferromagnetic single-electron
transistor in a three terminal configuration has been realized and studied
theoretically.\cite{RefOno,RefBarnas}

If a normal metal (N) is attached to a superconductor (S), quasiparticles of
different spins are coupled via Andreev reflection on the normal side of the
NS interface.\cite{RefAndreev} A strongly modified density of states (DOS)
in the normal metal caused by induced superconducting correlations was
found. \cite{RefMcMillan} This is so-called superconducting proximity
effect. In contrast with normal metals, in a ferromagnet (F) the presence of
a strong exchange field leads to big differences between the two spin bands.
The question of the coexistence of ferromagnetism and superconductivity has
been extensively investigated in the last decades. The effect of spin
splitting on the superconducting proximity effect was already investigated
long time ago\cite{RefGallagher}. New experimental developments on proximity
effect in ballistic ferromagnetic layers have been recently reported \cite
{RefKontos} and theoretically confirmed.\cite{RefMalek} Results obtained in
ferromagnet-superconductor nanocontacts have been explained in terms of bias
dependent transparency of the FS interface.\cite{RefUpadhyay,RefJong}
Diffusive heterostructures of ferromagnets and superconductors showed
unexpected large values of the conductance of the ferromagnetic part.\cite
{RefPetrasov,RefGiroud,RefLawrence,RefAumentado} Strong mutual influence
between the superconductor and the ferromagnet and long range proximity
effects have been proposed to explain these results.\cite
{RefWolfgang,RefBergeret} Effects related to the interplay between spin
accumulation and Andreev reflection has been also investigated.\cite
{RefFalko} On the other hand, the prediction of an exotic superconducting
state formed in a superconductor by the presence of an exchange field $h$,
was independently reported by Larkin-Ovchinnikov and Fulde-Ferrel, some
years ago.\cite{RefLarkin} Experimental confirmation of these effect in bulk
superconductors is still needed. Many investigations have focused on the
study of thermodynamic properties of FS multilayers.\cite
{RefRadovic,RefJiang,RefAarts,RefRyazanov}

Non-equilibrium Keldysh Green's functions in the quasiclassical
approximation, have been extensively used in the past to describe
non-equilibrium superconductivity.\cite{RefEilenberger,RefWolfgang1} This 
{\em quasiclassical theory of superconductivity} is based on semiclassical
transport equations for quasiparticles. Generally to solve these transport
equations is technically difficult, however the final results are relative
simple and clear. The {\em so-called} \ ``circuit theory of Andreev
reflection''\cite{RefYuli}, was conceived as a generalization of the
Kirchhoff's theory of electronic circuits, to simplify these equations into
a handful of accessible rules. The circuit theory of mesoscopic transport
was recently extended to describe transport in non-collinear magnetic
structures \cite{RefCircuit}. Interesting phenomena like spin precession
effects on the induced spin accumulation have been found \cite{RefDani}. An
important concept in the theory of non-collinear spin transport is the
so-called {\em mixing conductance }$G^{\uparrow \downarrow }$ \cite
{RefCircuit} which is closely related to the spin-current induced
magnetization torques \cite{RefGerrittrans} and the Gilbert damping in thin
ferromagnetic films.\cite{RefYaroslav} Moreover a three terminal
spin-transistor device has been proposed in the framework of the circuit
theory. \cite{RefCircuit,RefGerrittrans} The main advantage of the circuit
theory description is that it provides a simple approach based on
spin-charge current conservation to calculate the transport properties of
(multi-terminal) mesoscopic hybrid systems. This is achieved by mapping the
concrete geometry onto a topologically equivalent circuit, represented by
finite elements. We note that this very same description allows to calculate
a variety of transport properties, such as current statistics\cite
{yuli:99,wn01}, weak localization corrections\cite{yuli:99} or transmission
eigenvalues\cite{yulixx}. Consequently, it is worthwhile to extend this
general framework to the all possible combinations of heterostructures.

So far, the developed circuit theory is suitable to describe electron and
spin transport in multiterminal hybrid ferromagnet-normal metal FN {\em or}
superconductor-normal metal SN systems. Obviously, an extension of the
circuit theory to hybrid systems combining ferromagnets, normal metals {\em %
and} superconductors is necessary. This is done in the present paper. We
derive a general expression for an arbitrary contact, which, however, turns
out to be a little bit unhandy. To obtain manageable expressions we present
also approximate results for two special cases. Below we list all our
results:

\begin{itemize}
\item  The general matrix current Eq.(\ref{GeneralBC}) through an arbitrary
spin-dependent connector. It requires the knowledge of the full scattering
matrix (or transmission matrix) as consequence of the non-separable Spin-
and Nambu structures. This inhibits the transformation into the normal
eigenmodes of the scattering problem. Eq.~(\ref{GeneralBC}) is therefore
mostly of numerical interest.

\item  The matrix current Eq.(\ref{I_1_final}) for a weakly spin-dependent
contact. Here an expansion in terms of normal eigenmodes is possible. Eq.~(%
\ref{I_1_final}) is a spin-dependent correction to the matrix current from
Ref.~(\onlinecite{RefYuli}).

\item  If the spin-dependent contact is a tunnel barrier, the tunneling
matrix current Eq.~(\ref{Iprl}) takes a particular simple and transparent
expression. The properties of the contact can be expressed in terms of the
spin conductances $G_{\uparrow (\downarrow )}$ and the (complex) mixing
conductance $G^{\uparrow \downarrow }$\cite{RefCircuit}.
\end{itemize}

The paper is organized as follows. In section \ref{circuittheory} we
illustrate some basic aspects of the circuit theory. In Section \ref
{arbitraryconnector} we present the microscopic description of a contact
region or {\em connector} in the circuit theory. In section \ref
{generalmatrixcurrent} we calculate the matrix current through a general
connector between two metallic regions (nodes) of the circuit. The nodes can
be of N-, F- or S-type and the contact is assumed to have a arbitrary
magnetic structure ({\em e.~g.} magnetic tunnels junctions, magnetic
interfaces). This expression constitutes the generalized matrix current for
the circuit theory. In section \ref{twospecialcases} we make use of a
perturbation expansion in the spin structure to obtain simplified
expressions for two cases. Details of the pertubation expansion are
presented in the Appendix. In Section \ref{conclusions} we present our
conclusions.

\section{CIRCUIT THEORY}

\label{circuittheory}

In the circuit theory the system is split up into reservoirs (voltage
sources), connectors (contacts,interfaces) and nodes (low resistance
islands/wires) in analogy to classical electric circuits. Both reservoirs
and nodes are characterized by $16N_{ch}\times 16N_{ch}$ Green's functions $%
\check{G}$, which are matrices in Keldysh$\otimes $Nambu$\otimes $Spin$%
\otimes $Channels space, where $N_{ch}$ are the number of propagating modes.
These Green's functions play the role of {\em generalized potentials }of the
circuit theory. The Green's functions in the reservoirs and nodes are
assumed to be isotropic in momentum space. This requires that sufficient
elastic scattering is present both in reservoirs and nodes, due to the
presence of random scatterers and irregularities in the shape. This
justifies the use of the diffusion approximation to describe transport. This
assumption is reasonable since hybrid (multi-terminal) devices are quite
dirty systems. In general for the stationary case, $\check{G}$ depends on
space coordinates and energy%
\begin{mathletters}%
%
\begin{equation}
\check{G}(\vec{r},\vec{r}^{\prime },\varepsilon )=\int dt\text{ }\check{G}(%
\vec{r},\vec{r}^{\prime },t-t^{\prime })\text{ }\exp \{i\frac{\varepsilon }{%
\hbar }(t-t^{\prime })\}  \label{defG1}
\end{equation}
being 
\begin{equation}
\check{G}(\vec{r},\vec{r}^{\prime },t-t^{\prime })=\left[ 
\begin{array}{cc}
\hat{G}^{R}(\vec{r},\vec{r}^{\prime },t-t^{\prime }) & \hat{G}^{K}(\vec{r},%
\vec{r}^{\prime },t-t^{\prime }) \\ 
0 & \hat{G}^{A}(\vec{r},\vec{r}^{\prime },t-t^{\prime })
\end{array}
\right] .  \label{defG2}
\end{equation}
\begin{equation}
\hat{G}^{R}(\vec{r},\vec{r}^{\prime },t-t^{\prime })=-i\theta (t-t^{\prime
})\left\langle \left\{ \hat{\Psi}(\vec{r}),\hat{\Psi}^{\dagger }(\vec{r}%
^{\prime })\right\} \right\rangle  \label{defG3}
\end{equation}
\begin{equation}
\hat{G}^{A}(\vec{r},\vec{r}^{\prime },t-t^{\prime })=i\theta (t^{\prime
}-t)\left\langle \left\{ \hat{\Psi}(\vec{r}),\hat{\Psi}^{\dagger }(\vec{r}%
^{\prime })\right\} \right\rangle  \label{defG4}
\end{equation}
\begin{equation}
\hat{G}^{K}(\vec{r},\vec{r}^{\prime },t-t^{\prime })=-i\left\langle \left[ 
\hat{\Psi}(\vec{r}),\hat{\Psi}^{\dagger }(\vec{r}^{\prime })\right]
\right\rangle .  \label{DefG5}
\end{equation}
\end{mathletters}%
%
where $\left[ ..,..\right] $ and $\left\{ ..,..\right\} $ denotes commutator
and anticommutator respectively{\bf .} Semiclassical and diffusive
approximations allow to obtain Green's function which depends only in one
spatial coordinate $\check{G}(\vec{r},\varepsilon )$.\cite{RefRammer} On the
nodes the Green's functions are also assumed to be spatially homogeneous,
depending only on energy $\check{G}(\varepsilon )$. This requires that the
resistance of the node is much smaller than the contacts resistances
connecting different nodes, which implies also that the current through the
device is controlled by the contacts resistances. Regions with spatially
dependent Green's functions ({\em e.g.} diffusive wires) are modelled by an
appropriate discretization. For example, a quasi one-dimensional diffusive
wire can be represented by a series of tunnel junctions and nodes. Internal
dynamics along the wire (finite energy transport, spin-flip, etc.) is
included in this description as a leakage current from the node.\cite
{RefYuli}

In analogy to Ohm's law in classical electric circuits ($I=V/R$), in the
present circuit theory it is essential to obtain the ``spin-charge'' matrix
current that flows between two nodes (or between a reservoir and a node)
through a contact/connector. This matrix current depends on the properties
of the connector (analog to the resistance $R$) and on the Green's functions
at both side of the connector (analog to the voltage drop through the
connector $V$ ). In general the matrix current $\check{I}(z,\varepsilon )$
is defined in Keldysh$\otimes $Nambu$\otimes $Spin$\otimes $Channels space
as 
\begin{equation}
\check{I}(z,\varepsilon )=\frac{e^{2}\hbar }{m}\int d\rho \left( \frac{%
\partial }{\partial z}-\frac{\partial }{\partial z^{\prime }}\right) \check{G%
}(\vec{r},\vec{r}^{\prime },\varepsilon )|_{\vec{r}\text{ }=\text{ }\vec{r}%
^{\prime }}  \label{current1}
\end{equation}
where $\vec{r}\equiv (\vec{\rho},z)$, $z$ being along the transport
direction and $\vec{\rho}$ being perpendicular to the transport direction.
The electric current $I_{e}$ is defined in terms of the matrix current $%
\check{I}(z,\varepsilon )$ as 
\[
I_{e}(z)=\frac{1}{\text{ }4\text{ }e}\int_{-\infty }^{\infty }\frac{%
d\varepsilon }{2\pi \hbar }\text{Tr}\left\{ \hat{\sigma}_{z}\hat{\tau}_{3}%
\text{ }\check{I}^{\text{K}}(z,\varepsilon )\right\} , 
\]
where $\hat{\vec{\sigma}}$, $\hat{\tau}$ are Pauli matrices in spin and
Nambu space. $\check{G}(\vec{r},\vec{r}^{\prime };\varepsilon )$ can be
expanded in transverse modes in the following way\cite{RefZaitsev} 
\begin{eqnarray}
\check{G}(\vec{r},\vec{r}^{\prime };\varepsilon ) &=&\sum_{nm,\sigma ,\sigma
^{\prime }}\check{G}_{n\sigma ,m\sigma ^{\prime }}^{s,\text{ }s^{\prime
}}(z,z^{\prime })\times  \label{Zaitsev} \\
&&\exp (i\sigma k_{n}^{s}z-i\sigma ^{\prime }k_{m}^{s^{\prime }}z^{\prime
})\chi _{n}^{s}(\rho )\chi _{m}^{s^{\prime }\ast }(\rho ^{\prime }). 
\nonumber
\end{eqnarray}
Here $n$ ($m$) label the propagating modes, $\sigma ,$ $\sigma ^{\prime
}=\pm 1$ are the direction of propagation, $s,s\prime $ $\equiv \left\{
\uparrow ,\downarrow \right\} $ are spin indices, $k_{n}^{s}$ is the
longitudinal momentum and $\chi _{n}^{s}(\rho )$ is the transverse wave
function. By using this representation, the current can be written like 
\begin{eqnarray}
\check{I}^{s,s^{\prime }}(z,\varepsilon ) &=&i\text{ }e^{2}\sum_{n\text{ };%
\text{ }\sigma ,\sigma ^{\prime }}(\sigma \upsilon _{n}^{s}+\sigma ^{\prime
}\upsilon _{m}^{s^{\prime }})\check{G}_{n\sigma ,m\sigma ^{\prime }}^{s,%
\text{ }s^{\prime }}(z,\varepsilon )  \label{current2} \\
&&\times \int d\rho \text{ }\chi _{n}^{s}(\rho ,z)\chi _{m}^{s^{\prime }\ast
}(\rho ,z).  \nonumber
\end{eqnarray}
Here $\upsilon _{n(m)}^{s(s^{\prime })}=\hbar $ $k_{n(m)}^{s(s^{\prime })}/m$
is the velocity in mode $n(m)$ with spin $s(s^{\prime })$. The transverse
wave functions $\chi _{n}^{s}(\rho ,z)$ are eigenfunctions of the Spin
operator $S=\hbar $ $\hat{\sigma}_{z}/2$ and are normalized in a way that
the carry unity flux current. In this case the current reduces to 
\begin{equation}
\check{I}^{s,s}(z,\varepsilon )=2\text{ }i\text{ }e^{2}\sum_{n\text{ };\text{
}\sigma }\sigma \text{ }\upsilon _{n}^{s}\text{ }\check{G}_{n\sigma ,n\sigma
}^{s,s}(z,\varepsilon )  \label{current2N}
\end{equation}
Note that the current $\check{I}^{s,s}(z,\varepsilon )$ in Eq.(\ref
{current2N}) is a matrix diagonal in spin space. In the case of a
ferromagnetic material Eq.(\ref{current2N}) is valid if the magnetization of
the ferromagnetic material is parallel to the spin quantization axis. In the
case of non-collinear transport where there are two different magnetizations
defined in the system $\vec{M}_{1}$ and $\vec{M}_{2}$, Eq. (\ref{current2N})
has to be properly transformed into the basis of the spin quantization axis 
\[
\check{I}^{s,s^{\prime }}=U\text{ }\check{I}^{s,s}U^{-1} 
\]
$U$ $\ $being the spin rotation matrix that transforms $\vec{M}_{1(2)}$ into
the spin quantization axis. Note that in this case the transformed matrix $%
\check{I}^{s,s^{\prime }}$can have general spin structure.

\section{ARBITRARY CONNECTOR}

\label{arbitraryconnector}The contacts or {\em connectors } between the
reservoirs and nodes are described in terms of the scattering formalism by a
transfer matrix $\bar{M}$. The goal is to express the matrix current given
by Eq.(\ref{current2}) in terms of isotropic quasiclassical Green's
functions $\check{G}_{1(2)}$ at both sides of the contact and
transmission/reflection coefficients that characterize the scattering in the
contact region. $\bar{M}$ is in general a $16N_{ch}\times 16N_{ch}$ matrix
in Keldysh$\otimes $Nambu$\otimes $Spin$\otimes $Channels space.
Particularly $\bar{M}$ is proportional to the unit matrix in Keldysh space
and diagonal in Nambu space. This is denoted by $\left( \text{ }\bar{}%
\mbox{
}\right) $.

The Green's functions $\check{G}\equiv \check{G}_{n\sigma ,m\sigma ^{\prime
}}^{s,\text{ }s^{\prime }}(z,z^{\prime })$ is not a continuous function at $%
z=z\prime $.\cite{RefZaitsev,RefYuli} The values of $\check{G}_{n\sigma
,m\sigma ^{\prime }}^{s,\text{ }s^{\prime }}(z,z^{\prime })$ for $%
z>z^{\prime }$ and $z<z\prime $ are matched at $z=z\prime $ by using the
following representation (for details see Ref.\onlinecite{RefZaitsev,RefYuli}%
) 
\begin{eqnarray}
2i\check{G}_{n\sigma ,m\sigma ^{\prime }}^{s,\text{ }s^{\prime
}}(z,z^{\prime }) &=&\frac{\check{g}_{n\sigma ,m\sigma ^{\prime }}^{s,\text{ 
}s^{\prime }}(z,z^{\prime })}{\sqrt{v_{n}v_{m}}}  \label{match} \\
&&+\frac{\sigma \delta _{\sigma \sigma ^{\prime }}\text{ }\delta _{nm}\text{
sign}(z-z^{\prime })}{v_{n}}\check{1}.  \nonumber
\end{eqnarray}
where the matrix $\tilde{g}(z,z^{\prime })\equiv \check{g}_{n\sigma ,m\sigma
^{\prime }}^{s,\text{ }s^{\prime }}(z,z^{\prime })$ is continuous at $%
z=z\prime $. Note that the matrix $\check{g}_{n\sigma ,m\sigma ^{\prime
}}^{s,\text{ }s^{\prime }}$ has non-trivial structure in channels space. We
call this function {\em ballistic Green's function}.{\em \ }At the contact
region $(z=0)$ the transfer matrix $\bar{M}\equiv \bar{M}_{n\sigma ,m\sigma
^{\prime }}^{s,\text{ }s^{\prime }}$ connects $\tilde{g}(z,z^{\prime })$ at
both sides (left $z,z\prime =0^{-}$ and right $z,z\prime =0^{+}$) 
\begin{equation}
\tilde{g}_{(2)}=\bar{M}\text{ }\tilde{g}_{(1)\text{ }}\bar{M}^{\dagger \text{
}}  \label{gtransfer}
\end{equation}
where $\tilde{g}_{(2)}\equiv \tilde{g}(z=0^{+},z^{\prime }=0^{+})$ and $%
\tilde{g}_{(1)}\equiv \tilde{g}(z=0^{-},z^{\prime }=0^{-})$. Note that in
Eq.(\ref{gtransfer}) there is implicit a sumation over Channel and Spin
indices.

An important assumption of this theory is the {\em isotropization}
assumption. The ballistic Green's function $\tilde{g}$ defined at both sides
of the contact region becomes isotropic in momentum space, when departing
from the contact region. Such isotropization happens in a region of the
order of the elastic mean free path $l_{\text{mfp}}$ but much smaller than
the spin diffusion length $l_{\text{sf}}=\sqrt{D\text{ }\tau _{\text{sf}}}$
and the coherence length of a superconductor $\xi _{\Delta }=\sqrt{D/\Delta }
$. In the {\em isotropization zone} the dominant contribution to the
self-energy is the elastic scattering, which implies sufficient disorder (or
chaotic scattering) in this region. The self-energy is then $\check{\Sigma}%
_{imp}^{1(2)}=-i$ $\check{G}_{1(2)}/2$ $\tau _{imp}$, $\check{G}_{1(2)}$
being the isotropic quasiclassical Green's function for the left $(1)$ and
for the right $(2)$ side of the contact region, which is proportional to the
unit matrix in Channels space and $\tau _{imp}$ is the impurity scattering
time. In order to assure that the $\check{G}_{n\sigma ,m\sigma ^{\prime
}}^{s,\text{ }s^{\prime }}(z,z^{\prime })$ does not diverge at $z(z^{\prime
})\rightarrow 0$, we have to impose the following conditions\cite{RefYuli}

\begin{equation}
\left( \bar{\Sigma}^{z}+\check{G}_{1}\right) \left( \bar{\Sigma}^{z}-\tilde{g%
}_{1}\right) =0  \label{Iso1}
\end{equation}

\begin{equation}
\left( \bar{\Sigma}^{z}+\tilde{g}_{1}\right) \left( \bar{\Sigma}^{z}-\check{G%
}_{1}\right) =0  \label{Iso2}
\end{equation}

\begin{equation}
\left( \bar{\Sigma}^{z}-\check{G}_{2}\right) \left( \bar{\Sigma}^{z}+\tilde{g%
}_{2}\right) =0  \label{Iso3}
\end{equation}

\begin{equation}
\left( \bar{\Sigma}^{z}-\tilde{g}_{2}\right) \left( \bar{\Sigma}^{z}+\check{G%
}_{2}\right) =0,  \label{Iso4}
\end{equation}
$\bar{\Sigma}^{z}=${\bf $\sigma $ $\delta _{\sigma \sigma ^{\prime }}$}
being the z-Pauli matrix in the direction of propagation ``sub-space'' $%
\left\{ \sigma \sigma ^{\prime }\right\} $ of the Channels space. An
important consequence of this approach is that after the isotropization
zone, the ballistic Green's function $\check{g}_{n\sigma ,m\sigma ^{\prime
}}^{s,\text{ }s^{\prime }}(z,z)$ equals the isotropic quasiclassical Green's
function $\check{G}_{1(2)}$.

\section{GENERALIZED MATRIX CURRENT}

\label{generalmatrixcurrent}In previous work,\cite{RefYuli} $\bar{M}$ and $%
\check{G}_{1(2)}$ commute because no spin structure was considered. Now due
to their general spin structure, $\bar{M}$ and $\check{G}$ do not commute.
We multiply Eq.~(\ref{Iso1}) and Eq.~(\ref{Iso3}) by $\bar{M}$ from the left
and by $\bar{M}^{\dagger }$ from the right. Adding both resulting equations,
using Eq.(\ref{gtransfer}) and using the normalization condition $\check{G}%
^{2}=\check{1}$, \cite{RefRammer} we get the following expression for $%
\tilde{g}_{1}$ 
\begin{equation}
\tilde{g}_{1}=\left( \check{1}+\check{G}_{1}\bar{M}^{\dagger }\text{ }\check{%
G}_{2}\text{ }\bar{M}\right) ^{-1}\left[ 2\check{G}_{1}+\left( \check{1}-%
\check{G}_{1}\bar{M}^{\dagger }\text{ }\check{G}_{2}\text{ }\bar{M}\right) 
\bar{\Sigma}^{z}\right] \text{.}  \label{g1}
\end{equation}
Note that in Eq.(\ref{g1}) all the structure in channels is contained in $%
\bar{M}$. From Eq.(\ref{current2N}), the matrix current $\check{I}$ is be
expressed in terms of $\tilde{g}_{1(2)}$ as\cite{RefYuli} 
\begin{equation}
\check{I}(\varepsilon )=2\text{ }i\text{ }e^{2}\sum_{n\text{ };\text{ }%
\sigma }\sigma \text{ }\upsilon _{n}\text{ }\check{G}_{n\sigma ,n\sigma
^{\prime }}^{s,\text{ }s^{\prime }}(\varepsilon )=e^{2}\text{\mbox{Tr}}%
_{n,\sigma }\left[ \bar{\Sigma}^{z}\tilde{g}_{1}\right] .
\label{ballisticcurrent}
\end{equation}
Note that the Green's function $\check{G}_{n\sigma ,n\sigma ^{\prime }}^{s,%
\text{ }s^{\prime }}$ is assumed to be spatially homogeneous, depending only
on energy $\check{G}(\varepsilon )$. In this case that current does also not
depend on position. By using the cyclic property of the trace, we find
finally for the matrix current $\check{I}(\varepsilon )$ 
\begin{eqnarray}
\check{I}(\varepsilon ) &=&e^{2}\text{\mbox{Tr}}_{n,\sigma }\{\left( \check{1%
}+\check{G}_{1}\bar{M}^{\dagger }\text{ }\check{G}_{2}\text{ }\bar{M}\right)
^{-1}\left( \check{1}-\check{G}_{1}\bar{M}^{\dagger }\text{ }\check{G}_{2}%
\text{ }\bar{M}\right)  \label{GeneralBC} \\
&&+2\left( \check{1}+\check{G}_{1}\bar{M}^{\dagger }\text{ }\check{G}_{2}%
\text{ }\bar{M}\right) ^{-1}\text{ }\bar{\Sigma}^{z}\text{ }\check{G}_{1}\}.
\nonumber
\end{eqnarray}
Eq.(\ref{GeneralBC}) is the most general expression for the current $\check{I%
}$ in terms of isotropic quasiclassical Green's functions $\check{G}_{1(2)}$
at both sides of the contact and transmission/reflection coefficients of the
contact region. Once such expression of $\check{I}(\varepsilon )$ is
obtained, we can apply the generalized Kirchhoff's rules, imposing that the
sum of all matrix currents into a node is zero. This completely determines
all properties of our circuit. Eq.(\ref{GeneralBC}) therefore completes our
task to find the generalized boundary condition for the circuit theory.
However, in this form a concrete implementation requires the knowledge of
the full transfer matrix (or equivalently of the scattering matrix). Usually
this information is not available for realistic interfaces and one tries to
reduce Eq.~(\ref{GeneralBC}) to simple expressions by some reasonable
assumptions. In the next section we will do this for two special cases.
Note, that for a spin-independent interface, Eq.~(\ref{GeneralBC}) can be
expressed by the transmission eigenvalues only, which is a formidable
simplification. This transformation is not possible anymore for the
spin-dependent contact.

\section{TWO SPECIAL CASES}

\label{twospecialcases}We want to obtain more transparent and clear
analytical expressions of Eq.(\ref{GeneralBC}). The price to pay for that is
loss of generality, since we have to make certain assumptions about the
contact. The main difficulty lies in the inversion of the matrix $\check{1}+%
\check{G}_{1}\bar{M}^{\dagger }\check{G}_{2}\bar{M}$ in Channel-Spin space.
Let us assume that the transfer matrix $\bar{M}$ can be split in the
following form 
\begin{equation}
\bar{M}=\bar{M}_{0}+\delta \bar{M}=\bar{M}_{0}(\check{1}+\delta \bar{X})\,,
\label{trail}
\end{equation}
where $\bar{M}_{0}$ is a transfer matrix with structure in channel space
only and proportional to the unit matrix in spin space, while $\delta \bar{M}%
\equiv \bar{M}_{0}\delta \bar{X}$ includes non-trivial structure in spin
space. Note that the matrices $\bar{M}_{0}$, $\check{G}_{1(2)}$ commute with
each other, whereas in general $\delta \bar{X}$ does not commute with $\bar{M%
}_{0}$ and $\check{G}_{1(2)}$. Assuming that $\delta \bar{X}\ll 1$, we can
perform a perturbation expansion of $\left( \check{1}+\check{G}_{1}\bar{M}%
^{\dagger }\text{ }\check{G}_{2}\text{ }\bar{M}\right) ^{-1}$ in the
parameter $\delta \bar{X}$ (see Appendix). The matrix $\ \bar{M}\equiv
\left( \bar{M}\right) _{n\sigma ,m\sigma ^{\prime }}^{s;s^{\prime }\text{ }}$
can now be diagonalized in the basis of eigenmodes $N(N):$ $\left( \bar{M}%
\right) _{n\sigma ,m\sigma ^{\prime }}^{s;s^{\prime }\text{ }}\rightarrow
\left( \bar{M}\right) _{N\sigma ,N\sigma ^{\prime }}^{s;s^{\prime }\text{ }}$%
.

To 0th order $\bar{M}\equiv \bar{M}_{0}$ commutes with $\check{G}_{1(2)}$.
In this case Eq.~(\ref{GeneralBC}) reduces to 
\begin{equation}
\check{I}^{(0)}(\varepsilon )=e^{2}\text{\mbox{Tr}}_{N,\sigma }\left[ \frac{%
\check{1}}{\check{1}+\bar{Q}_{0}\check{G}_{1}\check{G}_{2}}\left( 2\text{ }%
\bar{\Sigma}^{z}\text{ }\check{G}_{1}+\check{1}-\bar{Q}_{0}\check{G}_{1}%
\check{G}_{2}\right) \right] .  \label{Yulicurrent}
\end{equation}
The hermitian conjugate matrix $\bar{Q}_{0}$ ($\bar{Q}_{0}^{\dagger }\equiv 
\bar{Q}_{0}$), reads in Channel space\cite{RefStone} 
\begin{equation}
\bar{Q}_{0}=\left[ 
\begin{array}{cc}
A & B \\ 
B^{\dagger } & A
\end{array}
\right]  \label{Qo_old}
\end{equation}
being $A$ real and $B$ complex $N_{ch}\times N_{ch}$ matrices.

The eigenvalues of $\bar{Q}_{0}$, appear in inverse pairs $\left(
q_{N},q_{N}^{-1}\right) $, and are related with the transmission
coefficients $T_{N}$ in the following way\cite{RefStone} 
\[
A=\frac{q_{N}+q_{N}^{-1}}{2}=\frac{2-T_{N}}{T_{N}} 
\]
being 
\[
\left| B\right| ^{2}=A^{2}-1. 
\]
By performing the trace over directions of the mode indices $\sigma (\sigma
^{\prime })$, the expression for the 0th order current reduces to 
\begin{equation}
\check{I}^{(0)}(\varepsilon )\equiv e^{2}%
\mathop{\textstyle\sum}%
\limits_{N}2T_{N}\frac{\left[ \check{G}_{2},\check{G}_{1}\right] }{%
4+T_{N}(\left\{ \check{G}_{1},\check{G}_{2}\right\} -2)}.
\label{Yulicurrent_comutator}
\end{equation}
Eq.(\ref{Yulicurrent_comutator}) is the expression for the matrix current
obtained in Ref.(\onlinecite{RefYuli}) for a spin-independent contact.

\subsection{CASE I: WEAK FERROMAGNETIC CONTACT.}

Now we concentrate in the first order term $\check{I}^{(1)}(\varepsilon )$,
given by Eq.(\ref{matrixcurrentperturbation1}) of the Appendix, $\delta \bar{%
X}\equiv \left( \delta \bar{X}\right) _{A}$ being anti-symmetric with
respect to time-reversal transformation. In particular $\left( \delta \bar{X}%
\right) _{A}$ is anti-symmetric with respect to transformation over spin
space and symmetric with respect to transformation over directional space.
In this case $\left( \delta \bar{X}\right) _{A}$ describes the case of a
weak ferromagnetic contact $\left( \delta \bar{X}\right) _{A}\sim \left( 
\vec{M}\text{ }\hat{\vec{\sigma}}\text{ }\hat{\tau}_{3}\right) $. The unity
vector $\vec{M}$ is in the direction of the magnetization, and $\hat{\vec{%
\sigma}}$, $\hat{\tau}$ are Pauli matrices in spin and Nambu space,
respectively. On the other hand, the $\left( \delta \bar{X}\right) _{S}$ may
describe spin-flip processes due to spin-orbit interaction at the contact.
At first order in $\delta \bar{X}$ the contribution given by $\left( \delta 
\bar{X}\right) _{S}$ vanishes. In general, to threat spin-flip at the
contact, we need to go to higher orders in $\delta X$ $\left( \delta
X^{2},\delta X^{3},..\right) $. From Eq.(\ref{IdTd}), Eq.(\ref{d_-1}) and
Eq.(\ref{d_-2}) (see Appendix), $\check{I}_{1}$ can be written as 
\begin{eqnarray}
\check{I}^{(1)}(\varepsilon ) &=&e^{2}\sum_{N}\frac{2}{4+T_{N}(\left\{ 
\check{G}_{1},\check{G}_{2}\right\} -2)}\times  \nonumber \\
&&\left[ \left\{ t_{2}^{\dagger }\delta t_{2}+\delta t_{2}^{\dagger }t_{2},%
\check{G}_{2}\right\} +\check{C},\check{G}_{1}\right]  \label{I_1_final} \\
&&\times \frac{2}{4+T_{N}(\left\{ \check{G}_{1},\check{G}_{2}\right\} -2)} 
\nonumber
\end{eqnarray}
where 
\begin{equation}
\check{C}=(t_{2}^{\dagger }\delta t_{2}-\delta t_{2}^{\dagger }t_{2})(\frac{%
4-2T_{N}}{T_{N}})-4\left( \text{ }t_{2}^{\dagger }\delta t_{2}+r_{1}^{\text{ 
}\dagger }\delta r_{1}\right) .  \label{C}
\end{equation}
Eq.(\ref{I_1_final}) is the spin-dependent correction to the matrix current
given in Eq.(\ref{Yulicurrent_comutator}). This result constitutes an
important step in the applicability of the circuit theory for F%
\mbox{$\vert$}%
N%
\mbox{$\vert$}%
S systems, because it provides of a simple prescription to describe
magnetically active interfaces/contacts.

\subsection{CASE II: TUNNELING BARRIERS}

For the case when the contact are tunneling barriers $(T_{N}\ll 1),$ is
possible to neglect the term $T_{N}(\left\{ \check{G}_{1},\check{G}%
_{2}\right\} -2)$ in the denominators in Eq.(\ref{Yulicurrent_comutator})
and Eq.(\ref{I_1_final}). Keeping terms of order $T_{N}$, $t_{2}^{\dagger
}\delta t_{2}$/$T_{N}$, $t_{2}^{\dagger }\delta t_{2}$ and $r_{1}^{\text{ }%
\dagger }\delta r_{1}$, we can write the total matrix current $\check{I}%
(\varepsilon )$ in a very transparent way like%
\begin{mathletters}%
%
\begin{eqnarray}
\frac{\check{I}(\varepsilon )}{2\pi \hbar } &=&\frac{G_{\text{T }}}{2}\left[ 
\check{G}_{2},\check{G}_{1}\right] +\frac{G_{\text{MR }}}{4}\left[ \left\{ 
\vec{M}\text{ }\hat{\vec{\sigma}}\text{ }\hat{\tau}_{3},\check{G}%
_{2}\right\} ,\check{G}_{1}\right]  \label{Iprl} \\
&&+i\frac{G_{\phi }}{2}\left[ \vec{M}\text{ }\hat{\vec{\sigma}}\text{ }\hat{%
\tau}_{3},\check{G}_{1}\right] .  \nonumber
\end{eqnarray}
where 
\begin{eqnarray}
G_{\text{T }} &=&G_{Q}\sum_{N}T_{N}  \label{GT} \\
\text{ }G_{\text{MR }} &=&G_{Q}\sum_{N}\delta T_{N}\text{ }  \label{GMR} \\
iG_{\phi }/2 &=&G_{Q}\sum_{N}(t_{2}^{\dagger }\delta t_{2}+r_{1}^{\dagger
}\delta r_{1}+  \label{Gfi} \\
&&\left( \frac{T_{N}-2}{2T_{N}}\right) \left( t_{2}^{\dagger }\delta
t_{2}-\delta t_{2}^{\dagger \text{ }}t_{2}\right) )  \nonumber
\end{eqnarray}
\end{mathletters}%
%
$G_{Q}\equiv e^{2}/2\pi \hbar $ being the conductance quantum and where the
explicit form of $\delta \bar{X}=\delta X$ $\vec{M}$ $\hat{\vec{\sigma}}$ $%
\hat{\tau}_{3}$ is used. For a weak ferromagnetic contact, the elements $%
\delta t$ $(\delta r)$ introduced in Eqs.~(\ref{small_dev_r1})-(\ref
{small_dev_t2}) in the Appendix can be expressed in terms of spin dependent
amplitudes like 
\begin{equation}
\hat{t}=\left[ 
\begin{array}{cc}
t^{\uparrow \uparrow } & 0 \\ 
0 & t^{\downarrow \downarrow }
\end{array}
\right] =\left[ 
\begin{array}{cc}
t+\delta t & 0 \\ 
0 & t-\delta t
\end{array}
\right]  \label{matrixT}
\end{equation}
\begin{equation}
\hat{r}=\left[ 
\begin{array}{cc}
r^{\uparrow \uparrow } & 0 \\ 
0 & r^{\downarrow \downarrow }
\end{array}
\right] =\left[ 
\begin{array}{cc}
r+\delta r & 0 \\ 
0 & r-\delta r
\end{array}
\right]  \label{matrixR}
\end{equation}
From these it easy to see that 
\begin{equation}
G_{\text{T }}=\sum_{N}G_{Q\text{ }}T_{N}=\sum_{N}G_{Q\text{ }}\frac{%
T_{N}^{\uparrow }+T_{N}^{\downarrow }}{2}=\frac{G_{\text{T}}^{\uparrow }+G_{%
\text{T}}^{\downarrow }}{2}  \label{GT_up+Down}
\end{equation}
\begin{equation}
G_{\text{MR }}=\sum_{N}G_{Q\text{ }}\delta T_{N}=\sum_{N}G_{Q\text{ }}\frac{%
T_{N}^{\uparrow }-T_{N}^{\downarrow }}{2}=\frac{G_{\text{T}}^{\uparrow }-G_{%
\text{T}}^{\downarrow }}{2}.  \label{GMR_up_down}
\end{equation}
Now we see that $G_{\text{T }}$is the usual conductance of the contact $G_{%
\text{T }}=\left( G_{\text{T}}^{\uparrow }+G_{\text{T}}^{\downarrow }\right)
/2$, and $G_{\text{MR}}=\left( G_{\text{T}}^{\uparrow }-G_{\text{T}%
}^{\downarrow }\right) /2$ accounts for the different conductances for
different spin directions, leading to a spin polarized current through the
junction.

On the other hand, for the case $T_{N}=0$ (the case of a ferromagnetic
insulator contact), Eq.(\ref{Iprl}) is not zero and reduces to 
\begin{equation}
\check{I}=i\frac{G_{\phi }}{2}\left[ \vec{M}\text{ }\hat{\vec{\sigma}}\text{ 
}\hat{\tau}_{3},\check{G}_{1}\right] .  \label{I_h}
\end{equation}
where $G_{\phi }$ depends only on reflection amplitudes at the normal metal
side 
\begin{equation}
iG_{\phi }/2=G_{Q\text{ }}\sum_{N}\frac{\left( r_{1}^{\uparrow \uparrow
}\left( r_{1}^{\downarrow \downarrow }\right) ^{\dagger }-r_{1}^{\downarrow
\downarrow }\left( r_{1}^{\uparrow \uparrow }\right) ^{\dagger }\right) _{N}%
}{4}.  \label{Gfi_r}
\end{equation}
The physical meaning of this term in this particular case, can be now
understood as follows: electrons with different spin directions pick up
different phases when reflecting at the ferromagnetic insulator. During this
process, ferromagnetic correlations are induced in the normal metal node. In
this particular case, the coefficient $G_{\phi }$ is related to the mixing
conductance introduced in \cite{RefCircuit} via $G_{\phi }=\text{Im}%
G^{\uparrow \downarrow }$. Eq.(\ref{Iprl}) has been recently used as
boundary condition current for the circuit theory applied in a MI%
\mbox{$\vert$}%
N%
\mbox{$\vert$}%
S structure, being MI\ a magnetic insulator and under conditions of
superconducting proximity effect. In Ref.(\onlinecite{RefDaniPRL}) we shown
that in this case, the effect of the conductance $G_{\phi }$ can be seen as
an induced magnetic field which give rise to spin-splitting of the induced
``BCS like'' density of states $\tilde{h}\equiv G_{\phi }\delta /2G_{Q}$,
being $\delta $ the average level spacing in the normal node. In
particular,we also show that for a system composed by two coupled MI%
\mbox{$\vert$}%
N%
\mbox{$\vert$}%
F trilayer structures the {\em absolute spin-valve effect} can be achieved
for a finite range of voltages.\cite{RefDaniPRL}

\section{CONCLUSIONS}

\label{conclusions}The circuit theory of mesoscopic transport is a
systematic way to describe transport in multiterminal hybrid structures in
which general rules analog to the Kirchhoff's rules of classical circuits
are used to {\em solve the circuit} and compute the transport properties of
the system under study. One of these rules imposes that the sum of all
``matrix current'' into a node must be zero. That is why, the expression of
the ``spin-charge'' matrix current that flows between two nodes (or between
a reservoir and a node) through a contact/{\em connector} is as essential
for the {\em the circuit theory},\ as the Ohm's law is for classical
electric circuits. In this paper we have generalized the expression of such
matrix currents for the case of multiterminal systems that included
ferromagnetic and superconducting reservoirs connected through magnetically
active contacts to one or several normal nodes.

We have derived the most general expression for this matrix current Eq.(\ref
{GeneralBC}) in terms of of isotropic quasiclassical Green's functions $%
\check{G}_{1(2)}$ at both sides of the contact which describe the adjacent
reservoirs/nodes and transmission/reflection coefficients that characterize
the scattering in the contact region. This expressions should be numerically
implemented in order to solve any general arrangement of reservoirs, contact
and nodes.

Moreover, we have perform a perturbation expansion in the spin asymmetry of
the transfer matrix $\delta \bar{X}$ associated to the contact region in
order to gain more knowledge and obtain more transparent expressions. We
found the expression for the matrix current that describes a weak
ferromagnetic contact to first order in $\delta \bar{X}$ (Eq.(\ref{I_1_final}%
)). In order to describe spin-flip processes at the contacts we need to go
to higher orders in the asymmetry $\delta \bar{X}$. For the case of a tunnel
barrier, the tunneling current takes a very simple and clear form (Eq.(\ref
{Iprl})). This expression is characterized by three conductance parameters $%
G_{\text{T }}$,$G_{\text{MR }}$and $G_{\phi }$, which can be expressed in
terms of the spin conductances $G_{\uparrow (\downarrow )}$ and the
(complex) mixing conductance $G^{\uparrow \downarrow }$.

We thank discussions with Ya. M. Blanter, A. Brataas, and Gerrit E. W.
Bauer. This work was financially supported by the Stichting voor
Fundamenteel Onderzoek der Materie (FOM).

\appendix
%

\section{PERTURBATION EXPANSION}

We perform a perturbation expansion of $\left( \check{1}+\check{G}_{1}\bar{M}%
^{\dagger }\text{ }\check{G}_{2}\text{ }\bar{M}\right) ^{-1}$ in terms of
the parameter $\delta \bar{X}$ as 
\begin{mathletters}%
%
\begin{equation}
\left( \check{1}+\check{G}_{1}\bar{M}^{\dagger }\text{ }\check{G}_{2}\text{ }%
\bar{M}\right) ^{-1}=\frac{\check{1}}{\check{1}+\check{A}}-\frac{\check{1}}{%
\check{1}+\check{A}}\delta \bar{M}^{(1)}\frac{\check{1}}{\check{1}+\check{A}}%
+..  \nonumber
\end{equation}

where 
\begin{equation}
\bar{Q}_{0}=\bar{M}_{0}^{\dagger }\text{ }\bar{M}_{0}  \label{PERT1}
\end{equation}
\begin{equation}
\check{A}=\bar{Q}_{0}\check{G}_{1}\check{G}_{2}  \label{PERT2}
\end{equation}
\begin{equation}
\delta \bar{M}^{(1)}=\check{G}_{1}\check{G}_{2}\bar{Q}_{0}\delta \bar{X}+%
\check{G}_{1}\delta \bar{X}^{\dagger }\bar{Q}_{0}\check{G}_{2}  \label{PERT4}
\end{equation}
\end{mathletters}%
%

To first order in $\delta \bar{X}$ an expansion in terms of normal
eigenmodes is possible{\bf . }The matrix $\ \bar{M}\equiv \left( \bar{M}%
\right) _{n\sigma ,m\sigma ^{\prime }}^{s;s^{\prime }\text{ }}$ can be
diagonalized in the basis of eigenmodes $N(N)$. For each eigenmode $N$, the
elements of the transfer matrix $\left( \bar{M}\right) _{\sigma ,\sigma
^{\prime }}^{\text{ }s;s^{\prime }}$ can be expressed in terms of
spin-dependent transmission and reflection amplitudes at each side of the
junction as%
\begin{mathletters}%
%
\begin{equation}
m_{\text{ }+,+}^{s;s^{\prime }}=t_{2}^{s,s^{\prime }}-\sum_{\alpha \beta
}r_{1}^{s,\alpha }\left( t_{1}^{\alpha ,\beta }\right) ^{-1}r_{2}^{\beta
,s^{\prime }}  \label{M++}
\end{equation}
\begin{equation}
m_{+,-}^{s;s^{\prime }\text{ }}=\sum_{\alpha }r_{1}^{s,\alpha }\left(
t_{1}^{\alpha ,s^{\prime }}\right) ^{-1}  \label{M+-}
\end{equation}
\begin{equation}
m_{-,+}^{\text{ }s;s^{\prime }}=-\sum_{\alpha }\left( t_{1}^{s,\alpha
}\right) ^{-1}r_{2}^{\alpha ,s^{\prime }}  \label{M-+}
\end{equation}
\begin{equation}
m_{-,-}^{s;s^{\prime }\text{ }}=\left( t_{1}^{s,s^{\prime }}\right) ^{-1}
\label{M--}
\end{equation}
\end{mathletters}%
%
,where $s,s\prime \left( \alpha ,\beta \right) $ $\equiv \left\{ \uparrow
,\downarrow \right\} $ are spin labels and $1(2)$ denotes left(right) side
of the contact region. Usually the transmission and reflection amplitudes
are introduced in terms of the scattering matrix $S$ 
\begin{equation}
\bar{S}=\left[ 
\begin{array}{cc}
r_{2}^{s\text{ }s^{^{\prime }}} & t_{1}^{s\text{ }s^{^{\prime }}} \\ 
t_{2}^{s\text{ }s^{^{\prime }}} & r_{1}^{s\text{ }s^{^{\prime }}}
\end{array}
\right] .  \label{Scattering_m}
\end{equation}
where in this case $t^{s\text{ }s^{^{\prime }}}$ and $r^{s\text{ }%
s^{^{\prime }}}$are matrices in Spin space. The scattering matrix $\bar{S}$
and the transfer matrix $\bar{M}$ are equivalent descriptions of a contact
region. Nevertheless the transfer matrix $\bar{M}$ obeys a {\em %
multiplicative} composite rule, whereas the scattering matrix obeys a more
complicated composition rule.\cite{RefStone,RefBeenakker}.

\ By writing in analogous way the other term in Eq.(\ref{GeneralBC}) $\left( 
\check{1}-\check{G}_{1}\bar{M}^{\dagger }\text{ }\check{G}_{2}\text{ }\bar{M}%
\right) $ and after some algebra, we obtain the following expression for the
matrix current $\check{I}$ to first order in $\delta \bar{X}$ 
\begin{mathletters}%
%
\begin{equation}
\check{I}(\varepsilon )=\check{I}^{(0)}(\varepsilon )+\check{I}%
^{(1)}(\varepsilon )  \label{matrixcurrentperturbation}
\end{equation}
where 
\begin{equation}
\check{I}^{(0)}(\varepsilon )=e^{2}\text{\mbox{Tr}}_{N,\sigma }\left\{ \frac{%
\check{1}}{\check{1}+\bar{Q}_{0}\check{G}_{1}\check{G}_{2}}\left( \check{1}+2%
\text{ }\bar{\Sigma}^{z}\text{ }\check{G}_{1}-\bar{Q}_{0}\check{G}_{1}\check{%
G}_{2}\right) \right\}  \label{matrixcurrentperturbation0}
\end{equation}
\begin{eqnarray}
\check{I}^{(1)}(\varepsilon ) &=&-2\text{ }e^{2}\text{\mbox{Tr}}_{N,\sigma
}\{\frac{\check{1}}{\check{1}+\bar{Q}_{0}\check{G}_{1}\check{G}_{2}}\check{0}%
\frac{\check{1}}{\check{1}+\bar{Q}_{0}\check{G}_{1}\check{G}_{2}}
\label{matrixcurrentperturbation1} \\
&&\times \left( \check{1}+\bar{\Sigma}^{z}\text{ }\check{G}_{1}\right) \} 
\nonumber
\end{eqnarray}
with 
\begin{equation}
\check{0}=\left( \check{G}_{1}\check{G}_{2}\bar{Q}_{0}\delta \bar{X}+\check{G%
}_{1}\delta \bar{X}^{\dagger }\bar{Q}_{0}\check{G}_{2}\right) .  \label{O}
\end{equation}
\end{mathletters}%
%

A general property of the transfer matrix is the flux conservation which can
be expressed as 
\begin{equation}
\bar{M}^{\dagger }\text{ }\bar{\Sigma}^{z}\text{ }\bar{M}=\bar{\Sigma}^{z}
\label{flux}
\end{equation}
Substituting Eq. (\ref{trail}) in Eq.(\ref{flux}) we get 
\begin{equation}
(\check{1}+\delta \bar{X}^{\dagger })\text{ }\bar{M}_{0}^{\dagger }\text{ }%
\bar{\Sigma}^{z}\text{ }\bar{M}_{0}\text{ }(\check{1}+\delta \bar{X})=\bar{%
\Sigma}^{z}  \label{generalcondition}
\end{equation}
The matrix $\bar{M}_{0\text{ }}$ should also obey flux conservation 
\begin{equation}
\bar{M}_{0}^{\dagger }\text{ }\bar{\Sigma}^{z}\text{ }\bar{M}_{0}=\bar{\Sigma%
}^{z}\text{,}  \label{fluxo}
\end{equation}
\ so we found the following equation for $\delta \bar{X}$ 
\begin{equation}
\delta \bar{X}^{\dagger }\bar{\Sigma}^{z}+\bar{\Sigma}^{z}\text{ }\delta 
\bar{X}+\delta \bar{X}^{\dagger }\bar{\Sigma}^{z}\text{ }\delta \bar{X}=0.
\label{condition}
\end{equation}
To first order in $\delta \bar{X}$ we have 
\begin{equation}
\delta \bar{X}^{\dagger }=-\bar{\Sigma}^{z}\text{ }\delta \bar{X}\text{ }%
\bar{\Sigma}^{z}.  \label{hermit_conjug_dX}
\end{equation}

From Eq.(\ref{hermit_conjug_dX}) we obtain for the elements of $\delta \bar{X%
}$%
\begin{mathletters}%
%
\begin{equation}
\delta \bar{X}_{+\text{ }+}^{\dagger }=-\delta \bar{X}_{+\text{ }+}
\label{dX11}
\end{equation}
\begin{equation}
\delta \bar{X}_{+\text{ }-}^{\dagger }=\delta \bar{X}_{-\text{ }+}
\label{dX12}
\end{equation}
\begin{equation}
\delta \bar{X}_{-\text{ }-}^{\dagger }=-\delta \bar{X}_{-\text{ }-}.
\label{dX22}
\end{equation}

\end{mathletters}%
%
We see that $\delta \bar{X}_{+\text{ }+}$ and $\delta \bar{X}_{-\text{ }-}$
are pure complex elements and $\delta \bar{X}_{+\text{ }-}$ and $\delta \bar{%
X}_{-\text{ }+}$ are complex conjugate.

The matrix $\bar{M}_{0}$ 
\begin{equation}
\bar{M}_{0}=\left[ 
\begin{array}{cc}
m_{+\text{ }+}^{0} & m_{+\text{ }-}^{0} \\ 
m_{-\text{ }+}^{0} & m_{-\text{ }-}^{0}
\end{array}
\right]  \label{Mo}
\end{equation}
is symmetric under time-reversal transformation, which involves both
transformation in Spin and Channels space, under the following prescription 
\begin{equation}
\widetilde{\bar{M}_{0}}=\vec{\sigma}^{y}\text{ }\bar{\Sigma}^{x}\text{ }%
\left( \bar{M}_{0}\right) ^{\ast }\text{ }\bar{\Sigma}^{x}\text{ }\vec{\sigma%
}^{y}=\bar{M}_{0}
\end{equation}
being $\bar{\Sigma}^{x}$ the x-Pauli matrix in directional space and $\vec{%
\sigma}^{y}$ the y-Pauli matrix in Spin space. As a result, the components
of $\bar{M}_{0}$ fulfill the following relations%
\begin{mathletters}%
%
\begin{equation}
m_{\text{ }+\text{ }+}^{0\text{ }\dagger }=m_{\text{ }-\text{ }-}^{0}
\label{mo++}
\end{equation}
\begin{equation}
m_{\text{ }+\text{ }-}^{0\text{ }\dagger }=m_{\text{ }-\text{ }+}^{0}
\label{mo+-}
\end{equation}
\end{mathletters}%
%

In general $\delta \bar{X}$ can be split in symmetric and anti-symmetric
parts with respect to time-reversal%
\begin{mathletters}%
%
\begin{equation}
\delta \bar{X}=\left( \delta \bar{X}\right) _{A}+\left( \delta \bar{X}%
\right) _{S}  \label{dXAS}
\end{equation}
\begin{equation}
\left( \widetilde{\delta \bar{X}}\right) _{A}=\vec{\sigma}^{y}\bar{\Sigma}%
^{x}\left( \delta \bar{X}\right) ^{\ast }\bar{\Sigma}^{x}\vec{\sigma}^{y}=-%
\text{ }\left( \delta \bar{X}\right) _{A}  \label{dXA}
\end{equation}
\begin{equation}
\left( \widetilde{\delta \bar{X}}\right) _{S}=\vec{\sigma}^{y}\bar{\Sigma}%
^{x}\left( \delta \bar{X}\right) ^{\ast }\bar{\Sigma}^{x}\vec{\sigma}%
^{y}=\left( \delta \bar{X}\right) _{S}  \label{dXS}
\end{equation}
\end{mathletters}%
%
Time-reversal transformation over Spin gives always an anti-symmetric
contribution $\widetilde{\delta \bar{X}}=\vec{\sigma}^{y}\left( \delta \bar{X%
}\right) ^{\ast }\vec{\sigma}^{y}=-$ $\delta \bar{X}$, so $\left( \widetilde{%
\delta \bar{X}}\right) _{A(S)}$ will be symmetric (anti-symmetric) with
respect to transformation in directional space $\left( \widetilde{\delta 
\bar{X}}\right) _{A(S)}=\bar{\Sigma}^{x}\left( \delta \bar{X}\right) ^{\ast }%
\bar{\Sigma}^{x}=\pm \left( \delta \bar{X}\right) _{A(S)}$. Taking into
account the structure of $\delta \bar{X}$ given by the condition $\delta 
\bar{X}^{\dagger }=-\bar{\Sigma}^{z}\delta \bar{X}$ $\bar{\Sigma}^{z},$ $%
\widetilde{\delta \bar{X}}$%
\begin{mathletters}%
%
\begin{equation}
\widetilde{\delta \bar{X}}=\left[ 
\begin{array}{cc}
-\delta X_{-\text{ }-} & \delta X_{+\text{ }-} \\ 
\delta X_{+\text{ }-}^{\ast } & -\delta X_{+\text{ }+}
\end{array}
\right] .  \label{dX_0}
\end{equation}
we find for $\left( \widetilde{\delta \bar{X}}\right) _{A}=\bar{\Sigma}%
^{x}\left( \delta \bar{X}\right) ^{\ast }\bar{\Sigma}^{x}=+\left( \delta 
\bar{X}\right) _{A}$, 
\begin{equation}
\widetilde{\delta \bar{X}}=\left[ 
\begin{array}{cc}
-\delta X_{-\text{ }-} & \delta X_{+\text{ }-} \\ 
\delta X_{+\text{ }-}^{\ast } & -\delta X_{+\text{ }+}
\end{array}
\right] =\left[ 
\begin{array}{cc}
\delta X_{+\text{ }+} & \delta X_{+\text{ }-} \\ 
\delta X_{+\text{ }-}^{\ast } & \delta X_{-\text{ }-}
\end{array}
\right] .  \label{dX_1}
\end{equation}
In this case $\delta X_{+\text{ }+}=-\delta X_{-\text{ }-}$ and $\delta \bar{%
X}_{A}$ is in directional Space 
\begin{equation}
\delta \bar{X}_{A}=\left[ 
\begin{array}{cc}
\delta X_{+\text{ }+} & \delta X_{+\text{ }-} \\ 
\delta X_{+\text{ }-}^{\ast } & -\delta X_{+\text{ }+}
\end{array}
\right] \sim \bar{\Sigma}^{z},\bar{\Sigma}^{x},\text{ }\bar{\Sigma}^{y}.
\label{dX_2}
\end{equation}
On the other hand for $\delta \bar{X}_{S}$, $\left( \widetilde{\delta \bar{X}%
}\right) _{S}=\bar{\Sigma}^{x}\left( \delta \bar{X}\right) ^{\ast }\bar{%
\Sigma}^{x}=-\left( \delta \bar{X}\right) _{S}$

\begin{equation}
\widetilde{\delta \bar{X}}=\left[ 
\begin{array}{cc}
-\delta X_{-\text{ }-} & \delta X_{+\text{ }-} \\ 
\delta X_{+\text{ }-}^{\ast } & -\delta X_{+\text{ }+}
\end{array}
\right] =\left[ 
\begin{array}{cc}
-\delta X_{+\text{ }+} & -\delta X_{+\text{ }-} \\ 
-\delta X_{+\text{ }-}^{\ast } & -\delta X_{-\text{ }-}
\end{array}
\right] .  \label{dX_3}
\end{equation}
Now $\delta X_{+\text{ }+}=-\delta X_{-\text{ }-}$, and $\delta X_{+\text{ }%
-}$ must to be zero. As a result $\delta \bar{X}_{S}$ is in directional
Space 
\begin{equation}
\delta \bar{X}_{S}=\left[ 
\begin{array}{cc}
\delta X_{+\text{ }+} & 0 \\ 
0 & \delta X_{+\text{ }+}
\end{array}
\right] \sim \bar{1}.  \label{dX_4}
\end{equation}

\end{mathletters}%
%
For the matrix $\delta \bar{M}=\bar{M}_{0}\delta \bar{X}$ we have 
\begin{eqnarray}
\delta M &=&\left[ 
\begin{array}{cc}
\delta m_{+\text{ }+} & \delta m_{+\text{ }-} \\ 
\delta m_{-\text{ }+} & \delta m_{-\text{ }-}
\end{array}
\right]  \label{dM} \\
&=&\left[ 
\begin{array}{cc}
\sum_{\sigma }m_{+\text{ }\sigma }\delta \bar{X}_{\sigma \text{ }+} & 
\sum_{\sigma }m_{+\text{ }\sigma }\delta \bar{X}_{\sigma \text{ }-} \\ 
\begin{array}{c}
\\ 
\sum_{\sigma }m_{-\text{ }\sigma }\delta \bar{X}_{\sigma \text{ }+}
\end{array}
& 
\begin{array}{c}
\\ 
\sum_{\sigma }m_{-\text{ }\sigma }\delta \bar{X}_{\sigma \text{ }-}
\end{array}
\end{array}
\right]  \nonumber
\end{eqnarray}
where $\sigma \equiv \left\{ +,-\right\} $. Analogy to $\bar{M}$, the matrix 
$\bar{S}$ can be written 
\begin{equation}
\bar{S}=\bar{S}_{0}+\delta \bar{S}  \label{SPERT}
\end{equation}
where $\bar{S}_{0}$ corresponds to the spin-independent part and $\delta 
\bar{S}$ accounts for the spin structure of $\bar{S}$. Now we assume that $%
\delta \bar{S}$ is a small deviation with respect to $\bar{S}_{0}$ $(\delta 
\bar{S}\ll \bar{S}_{0}).$ In this case the elements of $\bar{S}$ are 
\begin{mathletters}%
%
\begin{equation}
r_{1}^{s\text{ }s^{^{\prime }}}=r_{1}+\delta r_{1}  \label{small_dev_r1}
\end{equation}
\begin{equation}
t_{1}^{s\text{ }s^{^{\prime }}}=t_{1}+\delta t_{1}  \label{small_dev_t1}
\end{equation}
\begin{equation}
r_{2}^{s\text{ }s^{^{\prime }}}=r_{2}+\delta r_{2}  \label{small_dev_r2}
\end{equation}
\begin{equation}
t_{2}^{s\text{ }s^{^{\prime }}}=t_{2}+\delta t_{2}.  \label{small_dev_t2}
\end{equation}
\end{mathletters}%
%
The quantities $t_{1(2)}\left( r_{1(2)}\right) $ are spin-independent
transmission(reflection) amplitudes, whereas $\delta t_{1(2)}^{s\text{ }%
s^{^{\prime }}}\left( \delta r_{1(2)}^{s\text{ }s^{^{\prime }}}\right) $ are
the spin-dependent correction to the total transmission(reflection) $t_{1}^{s%
\text{ }s^{^{\prime }}}\left( r_{2}^{s\text{ }s^{^{\prime }}}\right) $
amplitudes. The elements of $\bar{S}$ can be expressed in terms of the
elements of $\bar{M}$ like%
\begin{mathletters}%
%
\begin{equation}
r_{2}^{s\text{ }s^{^{\prime }}}=-(m_{\text{ }-\text{ }-}^{s;s^{\prime
}})^{-1}\text{ }m_{\text{ }-\text{ }+}^{s;s^{\prime }}  \label{r1ss}
\end{equation}
\begin{equation}
t_{2}^{s\text{ }s^{^{\prime }}}=m_{+\text{ }+}^{\text{ }s;s^{\prime }}-m_{+%
\text{ }-}^{\text{ }s;s^{\prime }}(m_{-\text{ }-}^{\text{ }s;s^{\prime
}})^{-1}\text{ }m_{-\text{ }+}^{\text{ }s;s^{\prime }}  \label{t1ss}
\end{equation}
\begin{equation}
r_{1}^{s\text{ }s^{^{\prime }}}=m_{+\text{ }-}^{\text{ }s;s^{\prime }}(m_{%
\text{ }-\text{ }-^{\prime }}^{s;s})^{-1}  \label{r2ss}
\end{equation}
\begin{equation}
t_{1}^{s\text{ }s^{^{\prime }}}=(m_{-\text{ }-}^{\text{ }s;s^{\prime
}})^{-1}.\text{ }  \label{t2ss}
\end{equation}

\end{mathletters}%
%
By substituting Eq.(\ref{dM}) and Eqs.(\ref{small_dev_r1}-\ref{small_dev_t2}%
), using analog relations between the elements of $\bar{S}_{0}$ and $\bar{M}%
_{0}$ and by expanding $(m_{s;s^{\prime }}^{\text{ }-\text{ }-})^{-1}$ up to
order $\delta m_{-\text{ }-}$, we obtain%
\begin{mathletters}%
%
\begin{equation}
\delta r_{2}^{s\text{ }s^{^{\prime }}}=r_{2}\text{ }\delta X_{+\text{ }+}^{%
\text{ }s\text{ }s^{^{\prime }}}-\delta X_{-\text{ }+}^{\text{ }s\text{ }%
s^{\prime }}+(r_{2}\text{ }\delta X_{+\text{ }-}^{\text{ }s\text{ }%
s^{^{\prime }}}-\delta X_{-\text{ }-}^{\text{ }s\text{ }s^{^{\prime }}})%
\text{ }r_{2}  \label{dr1}
\end{equation}
\begin{equation}
\delta t_{2}^{s\text{ }s^{^{\prime }}}=t_{2}(\text{ }\delta X_{+\text{ }+}^{%
\text{ }s\text{ }s^{^{\prime }}}+r_{2}\text{ }\delta X_{+\text{ }-}^{\text{ }%
s\text{ }s^{^{\prime }}})  \label{dt1}
\end{equation}
\begin{equation}
\delta r_{1}^{s\text{ }s^{^{\prime }}}=t_{2}\text{ }t_{1}\text{ }\delta X_{+%
\text{ }-}^{\text{ }s\text{ }s^{^{\prime }}}  \label{dr2}
\end{equation}
\begin{equation}
\delta t_{1}^{s\text{ }s^{^{\prime }}}=t_{1}(r_{2}\text{ }\delta X_{+\text{ }%
-}^{\text{ }s\text{ }s^{^{\prime }}}-\delta X_{-\text{ }-}^{\text{ }s\text{ }%
s^{^{\prime }}}).  \label{dt2}
\end{equation}
\end{mathletters}%
%
So we see that the elements of $\delta \bar{X}$ can be expressed in terms of
the transmission and reflection amplitudes $\delta t$ $(\delta r)$. Note
that in Eq.(\ref{dr1})-Eq.(\ref{dt2}) we explicitly include the spin indices 
$s,s\prime $ of the elements $\delta \bar{X}$ and $\delta t$ $(\delta r)$ to
emphasize that both $\delta \bar{X}$ and $\delta t$ $(\delta r)$, contain
information about the spin structure of the contact region.

Now we consider the case of $\delta \bar{X}_{A}$ which corresponds to a weak
ferromagnetic contact $\left( \left( \delta \bar{X}\right) _{A}\sim \left( 
\vec{M}\text{ }\hat{\vec{\sigma}}\text{ }\hat{\tau}_{3}\right) \right) $. To
evaluate the trace of Eq. (\ref{matrixcurrentperturbation1}) we can re-write
the expression in the following way 
\begin{eqnarray}
\check{I}^{(1)}(\varepsilon ) &=&e^{2}\text{\mbox{Tr}}_{N,\sigma }\{-2\frac{%
T_{N}}{4+T_{N}(\left\{ \check{G}_{1},\check{G}_{2}\right\} -2)}\times
\label{I1T/} \\
&&\left( \check{G}_{2}\check{G}_{1}+\bar{Q}_{0}^{-1}\right) \left( \check{G}%
_{1}\check{G}_{2}\bar{Q}_{0}\delta \bar{X}+\check{G}_{1}\delta \bar{X}%
^{\dagger }\bar{Q}_{0}\check{G}_{2}\right)  \nonumber \\
&&\left( \check{G}_{2}\check{G}_{1}+\bar{Q}_{0}^{-1}\right) \left( \check{1}+%
\bar{\Sigma}^{z}\text{ }\check{G}_{1}\right)  \nonumber \\
&&\times \frac{T_{N}}{4+T_{N}(\left\{ \check{G}_{1},\check{G}_{2}\right\} -2)%
}\}  \nonumber
\end{eqnarray}

\bigskip The central term can be separate in four different terms as%
\begin{mathletters}%
%
\begin{equation}
\left( 
\begin{array}{c}
\bar{Q}_{0}\delta \bar{X}\text{ }\check{G}_{2}\check{G}_{1}+\check{G}%
_{2}\delta \bar{X}^{\dagger }\bar{Q}_{0}\check{G}_{1}+\check{G}_{1}\check{G}%
_{2}\text{ }\delta \bar{X}\text{ }\bar{Q}_{0}^{-1} \\ 
+\check{G}_{1}\text{ }\bar{Q}_{0}^{-1}\delta \bar{X}^{\dagger }\text{ }%
\check{G}_{2}
\end{array}
\right) +  \label{term1}
\end{equation}

\begin{equation}
\left( 
\begin{array}{c}
\bar{Q}_{0}\delta \bar{X}\bar{Q}_{0}^{-1}+\check{G}_{2}\delta \bar{X}%
^{\dagger }\check{G}_{2}+\check{G}_{1}\check{G}_{2}\text{ }\delta \bar{X}%
\text{ }\check{G}_{2}\check{G}_{1} \\ 
+\check{G}_{1}\text{ }\bar{Q}_{0}^{-1}\delta \bar{X}^{\dagger }\text{ }\bar{Q%
}_{0}\text{ }\check{G}_{1}
\end{array}
\right) +  \label{term2}
\end{equation}
\begin{equation}
\left( 
\begin{array}{c}
\bar{Q}_{0}\delta \bar{X}\text{ }\check{G}_{2}\check{G}_{1}+\check{G}%
_{2}\delta \bar{X}^{\dagger }\bar{Q}_{0}\check{G}_{1}+\check{G}_{1}\check{G}%
_{2}\text{ }\delta \bar{X}\text{ }\bar{Q}_{0}^{-1} \\ 
+\check{G}_{1}\text{ }\bar{Q}_{0}^{-1}\delta \bar{X}^{\dagger }\text{ }%
\check{G}_{2}
\end{array}
\right) \left( \bar{\Sigma}^{z}\text{ }\check{G}_{1}\right) +  \label{term3}
\end{equation}

\begin{equation}
\left( 
\begin{array}{c}
\bar{Q}_{0}\delta \bar{X}\bar{Q}_{0}^{-1}+\check{G}_{2}\delta \bar{X}%
^{\dagger }\check{G}_{2}+\check{G}_{1}\check{G}_{2}\text{ }\delta \bar{X}%
\text{ }\check{G}_{2}\check{G}_{1} \\ 
+\check{G}_{1}\text{ }\bar{Q}_{0}^{-1}\delta \bar{X}^{\dagger }\text{ }\bar{Q%
}_{0}\text{ }\check{G}_{1}
\end{array}
\right) \left( \bar{\Sigma}^{z}\text{ }\check{G}_{1}\right) .  \label{term4}
\end{equation}

\end{mathletters}%
%

The trace over ``direction'' indices $\sigma (\sigma ^{\prime })$ gives for
the first term Eq.(\ref{term1}) 
\begin{eqnarray}
&&2%
\mathop{\rm Re}%
[B^{\ast \text{ }}\delta X_{12}]\text{ }\check{G}_{2}\check{G}_{1}-\check{G}%
_{1}\check{G}_{2}\text{ }(2%
\mathop{\rm Re}%
[B^{\ast }\delta X_{12}])  \label{FIRSTERM} \\
&&+\check{G}_{2}\text{ }2%
\mathop{\rm Re}%
[B^{\ast \text{ }}\delta X_{12}]\text{ }\check{G}_{1}-\check{G}_{1}\text{ }2%
\mathop{\rm Re}%
[B^{\ast }\delta X_{12}]\text{ }\check{G}_{2}.  \nonumber
\end{eqnarray}
Note that $\check{G}_{1(2)}$ and $\delta X_{12}$ are still matrices in
``Keldysh-Nambu-spin'' space and do not commute with each other. $B^{\ast }$
can be written in terms of reflection and transmission amplitudes like 
\begin{equation}
B^{\ast }=2m_{+\text{ }+}m_{+\text{ }-}=-\frac{2}{T_{N}}r_{2}.  \label{B*}
\end{equation}
The spin-dependent corrections to the transmission and reflection
probabilities are defined like 
\begin{equation}
\delta T_{1(2)}=t_{1(2)\text{ }}^{\text{ }\dagger }\delta t_{1(2)}+\delta
t_{1(2)}^{\dagger \text{ }}t_{1(2)}  \label{dT}
\end{equation}
\begin{equation}
\delta R_{1(2)}=r_{1(2)\text{ }}^{\dagger }\delta r_{1(2)}+\delta
r_{1(2)}^{\dagger \text{ }}r_{1(2)}.  \label{dR}
\end{equation}
For this quantities we find 
\begin{equation}
\frac{\delta T_{1}}{T_{1}}\equiv \frac{\delta T_{2}}{T_{2}}\equiv -\frac{%
\delta R_{1}}{T_{1}}\equiv -\frac{\delta R_{2}}{T_{2}}=2%
\mathop{\rm Re}%
[r_{2}\text{ }\delta X_{+\text{ }-}^{\text{ }s\text{ }s^{^{\prime }}}].
\label{correction dT_dR}
\end{equation}
so 
\begin{equation}
2%
\mathop{\rm Re}%
[B^{\ast }\delta X_{+\text{ }-}^{\text{ }s\text{ }s^{\prime }}]=-\frac{2}{%
T_{N}}(2%
\mathop{\rm Re}%
[r_{2}\delta X_{+\text{ }-}^{\text{ }s\text{ }s^{\prime }}])=-\frac{2}{T_{N}}%
\frac{\delta T_{N}}{T_{N}}.  \label{2Re[]}
\end{equation}
At first order in $\delta \bar{X}$, the contribution given by Eq.(\ref{2Re[]}%
) is zero for in the case of $\left( \delta \bar{X}\right) _{S}$. Finally
The contribution to this first term for the $\check{I}_{1}$ current in terms
of $T_{N},$ $\delta T_{N}$ and $\check{G}_{1(2)}$ is 
\begin{eqnarray}
\check{I}^{(1)}(\varepsilon ) &=&e^{2}\sum_{N}\frac{2}{4+T_{N}(\left\{ 
\check{G}_{1},\check{G}_{2}\right\} -2)}\left[ \left\{ \delta \check{T}_{N},%
\check{G}_{2}\right\} ,\check{G}_{1}\right]  \label{FinlaI1} \\
&&\times \frac{2}{4+T_{N}(\left\{ \check{G}_{1},\check{G}_{2}\right\} -2)}+..
\nonumber
\end{eqnarray}
For the second term (Eq.\ref{term2}), using the property that the trace is
cyclic, it is easy to see that \mbox{Tr}$_{\sigma }\left\{ \delta \bar{X}%
\right\} \equiv $\mbox{Tr}$_{\sigma }\left\{ \delta \bar{X}^{\dagger
}\right\} \equiv $\mbox{Tr}$_{\sigma }\left\{ \bar{Q}_{0}\delta \bar{X}\bar{Q%
}_{0}^{-1}\right\} $

$\equiv $\mbox{Tr}$_{\sigma }\left\{ \bar{Q}_{0}^{-1}\delta \bar{X}^{\dagger
}\bar{Q}_{0}\right\} =0$, so the second term gives no contribution to $%
\check{I}_{1}$.

The third and fourth terms include the factor $\bar{\Sigma}^{z}$ $\check{G}%
_{1}$ in the right hand side. The presence of the matrix $\bar{\Sigma}^{z}$
will change the structure of this term in Channels space with respect to Eqs.%
\ref{term1} and \ref{term2}. Using that $\left[ \delta \bar{X},\check{G}_{2}%
\right] =0$, $\check{G}_{2}$ being the Green's function of a
(normal/ferromagnetic) reservoir, finally $\check{I}_{1}$ gives 
\begin{eqnarray}
\check{I}^{(1)}(\varepsilon ) &=&e^{2}\text{ }\sum_{N}\frac{2}{%
4+T_{N}(\left\{ \check{G}_{1},\check{G}_{2}\right\} -2)}\times  \nonumber \\
&&\left[ \left\{ \delta \check{T}_{N},\check{G}_{2}\right\} +\left(
4-2T_{N}\right) \text{ }\delta \check{\Theta}_{N}-4\text{ }\delta \check{\Xi}%
_{N}\text{ },\check{G}_{1}\right]  \label{IdTd} \\
&&\times \frac{2}{4+T_{N}(\left\{ \check{G}_{1},\check{G}_{2}\right\} -2)} 
\nonumber
\end{eqnarray}
being $\delta \check{\Xi}_{N}=T_{N}$ $\delta X_{+\text{ }+,N}$ and $\delta 
\check{\Theta}_{N}=2\left[ \delta X_{+\text{ }+}+i%
\mathop{\rm Im}%
\left( r_{2}\text{ }\delta X_{+\text{ }-}\right) \right] _{N}$ both pure
imaginary quantities. Finally inverting Eq.(\ref{dr1})-Eq.(\ref{dt2}) we can
express also the quantities $\delta \check{\Xi}_{N}$ and $\delta \check{%
\Theta}_{N}$ in terms of spin-independent transmission(reflection)
amplitudes $t_{1(2)}\left( r_{1(2)}\right) $,and its the spin-dependent
corrections $\delta t_{1(2)}\left( \delta r_{1(2)}\right) $%
\begin{equation}
\delta \check{\Xi}_{N}=t_{2}^{\dagger }\delta t_{2}+r_{1}^{\dagger }\delta
r_{1}  \label{d_-1}
\end{equation}
\begin{equation}
\delta \check{\Theta}_{N}=\frac{t_{2}^{\dagger }\delta t_{2}-\delta
t_{2}^{\dagger \text{ }}t_{2}}{T_{N}}  \label{d_-2}
\end{equation}

\end{multicols}%
%

\end{document}